\documentclass[aps,twocolumn]{revtex4}
\usepackage{graphicx,subfigure}
\usepackage{amsmath,amssymb}
\usepackage{bm}

\newcommand{\be}{\begin{eqnarray}}
\newcommand{\ee}{\end{eqnarray}}

\def\slashchar#1{\setbox0=\hbox{$#1$}           % set a box for #1 
   \dimen0=\wd0                                 % and get its size
   \setbox1=\hbox{/} \dimen1=\wd1               % get size of /
  \ifdim\dimen0>\dimen1                        % #1 is bigger
 \rlap{\hbox to \dimen0{\hfil/\hfil}}      % so center / in box
  #1                                        % and print #1
 \else                                        % / is bigger
    \rlap{\hbox to \dimen1{\hfil$#1$\hfil}}   % so center #1
    /                                         % and print /
 \fi}                                         %

\begin{document}

\title{  Correlations and fluctuations of the gauge topology  at finite temperatures}

\author{Rasmus Larsen and   Edward  Shuryak }

\affiliation{Department of Physics and Astronomy, Stony Brook University,
Stony Brook NY 11794-3800, USA}

\begin{abstract}
Instanton-dyons are topological solitons -- solutions of Yang-Mills equations --
which appear at non-trivial expectation value of $A_0$ at nonzero temperatures.
Using the ensembles of those, generated in our previous work, for 2-color and 2-flavor QCD,
below and above the deconfinement-chiral phase transition, we study the correlations
between them, as well as fluctuations of several global charges in the sub-volumes
of the total volume. The determined correlation lengths are the finite-$T$ 
extension of hadronic masses, such as that of $\eta'$ meson.
\end{abstract}
\maketitle

\section{Introduction}
\subsection{Instanton-dyon ensembles} 
Instantons, discovered in 1970's \cite{Belavin:1975fg} are 4-dimensional Euclidean topological solitons of the gauge theory. Ensembles of instantons were studied in 1980's and 1990's, in a frame of the so called ``instanton liquid model", for a review see \cite{Schafer:1996wv}.  They were shown to
explain explicit breaking of the $U(1)_a$  and spontaneous breaking of the $SU(N_f)$ chiral 
symmetries, and a large number of hadronic correlation functions.

The so called Polyakov line is used as a deconfinement order parameter, being nonzero at $T>T_c$.
Interpreting this as existence of nonzero average $A_0$ field, one needs to modify all classical solutions
respectively.  When such solutions were found in 1998 \cite{Kraan:1998sn,Lee:1998bb} it was realized
that 
instantons  get split into $N_c$ (number of colors) constituents, the selfdual {\em instanton-dyons}  connected only by (invisible) Dirac strings.
Since these objects have nonzero electric and magnetic charges and source
Abelian (diagonal) massless gluons, the corresponding ensemble is 
an ``instanton-dyon plasma", with long-range Coulomb-like forces between constituents.  

The first application of the instanton-dyons were made soon after their discovery
in the context of supersymmetric gluodynamics \cite{Davies:1999uw}. This paper 
set up an infinitely dilute but confining setting, and
solved a historically important puzzling
mismatch  of the value of the gluino condensate in this theory.
 Further work on semi-classical  confining regimes in parametrically dilute settings
  has been done by Poppitz, Unsal et al
~\cite{Poppitz:2011wy,Poppitz:2012sw} .

Recent progress  is related to studies of the 
semiclassical instanton-dyon ensembles.
The
 high-density confining phase can be studied analytically, in  the mean field
approximation \cite{Liu:2015ufa,Liu:2015jsa,Liu:2016thw,Liu:2016mrk,Liu:2016yij}.
Direct
numerical simulation of the dyon ensembles, started in \cite{Faccioli:2013ja},
have demonstrated that back reaction of the dyons on the holonomy potential
generates the deconfinement phase transition. It is 
of the second order for pure gauge $SU(2)$ theory
\cite{Larsen:2015vaa}, but it becomes a smooth cross over if two light quark flavors
are included 
\cite{Larsen:2015tso}.
The last theory also shows the chiral restoration phase transition, also a crossover, which occurs at approximately  the same temperature. Both phase transitions show strong changes \cite{Larsen:2016fvs}  as
a function of nontrivial quark periodicity phases (known also as flavor holonomies or imaginary
chemical potentials).  For a recent brief review
see \cite{Shuryak:2016vow}.

\subsection{Lattice studies}

Recent efforts has resulted in a substantial progress
in lattice evaluation of the topological susceptibility $\chi(T)$, and even higher moments of
the topological charge distribution,
in a wide range of  temperatures, see e.g.
\cite{Bonati,Borsanyi,Petreczky:2016vrs}.
Although there remain serious methodical questions, it is generally 
accepted that
at the high temperatures, say $T>2T_c$, these data can be explained by a dilute gas of
independent  instantons.  This statement holds even in QCD with (realistically) light  quarks,
where the instantons are highly suppressed by the product of quark masses. 

The main questions of the field are the following ones: 
 What are the main building blocks of the topological ensembles at lower  $T<2T_c$, 
 are they still the  {\em instantons}, or those get disassembled 
 into their constituents, the   {\em instanton-dyons}?  What is the
 density and manifestations of the neutral topological clusters, not contributing to $\chi$, such as the
 instanton-antiinstanton molecules?

We already mentioned the main lattice observable, the vacuum expectation value (VEV) of the Polyakov line. Its temperature dependence $<P(T)>$ 
 is by now well established.  In QCD it gradually changes  between zero and one when $T$ changes from $T_c$
to roughly $ 2.5T_c$. Lattice data on $\chi(T)$ in this region has not yet converged,
but it is already clear that it does not follow the dilute instanton gas power. 
The question remains whether indeed the topology 
ensemble can  be correctly described by a plasma of instanton constituents,
instanton dyons, and, if so, where and how this change in the basic topology units happens.

 The semiclassical theory of the instanton-dyons
 has  shown that their ensemble undergoes deconfinement and chiral
transitions near $T_c$. As we already mentioned, this theory
 semi-qualitatively reproduce the lattice results, both in pure glue and in QCD-like 
 settings.  The question now is how to make this comparison quantitative.
 In this respect  the ongoing efforts to
locate the instanton-dyons on the lattice, e.g.
 by Ilgenfritz and collaborators \cite{Bornyakov:2015xao}, with or without imaginary chemical
 potentials, are very important. 
 
 Another possible lattice tool, recently discussed by one of us \cite{Shuryak:2017fkh}, are the fluctuation of topological and magnetic charges on  sub-lattices. The present paper, in which we 
 calculate the correlations and fluctuations in our ensemble of the instanton-dyons,
 is another effort in this direction.

Finally, there remains a question of whether and how far can the instanton-dyon description  be extended into the confining regime, say to $T\sim 0.5 T_c$. At low $T\rightarrow 0$ the absolute magnitude of the holonomy
 field goes to zero, the KvBLL
phenomenon  disappears, and the Matsubara
 box becomes infinitely  large. So, perhaps 
 the relevant topological objects 
 return  to their 4-d symmetric form, the instantons.

\subsection{The main goals of this paper} \label{sec_goals}
 
 General questions outlined above can be put  
 in a more quantitative specific 
 form, provided one can compare certain observables measured on the lattice and
 in the semiclassical ensembles. 
 
 With this in mind, we provide a set of measured quantities, based on 
 instanton-dyon ensembles from Ref.\cite{Larsen:2015tso}. As emphasized above,
 those are expected to reproduce gauge topology in the temperature range 
  $0.5T_c<T<2 T_c$, but its boundaries are not yet well defined.
  Where exactly  the transition between such regimes of gauge topology take place
  will require more work. 
 
% It is generally recognized that in the absence of holonomy field, .  Therefore, 
%  at high temperatures, when $<P>\approx 1$,  
%  one should use the language of instantons. The instanton-dyon description should be 
%  adequate at the temperatures where $<P>$ changes from zero to one, that is 
%  for  
  %Another boundary of the  instanton-dyon description is 

The values  of various correlations lengths, if known,  provide important
dynamical insights.  The most discussed example is related with the $T=0$
correlator of the topological charges. The corresponding screening mass is
 that of the  $\eta'$ meson,  $$m_{\eta'}\approx 958\, MeV $$
The inverse of it, the screening length, is $\approx .2\, fm $, which is several times smaller than
the typical inter-soliton distances  $\approx 1\, fm $. What this phenomenological fact tells us is that the topological objects in the QCD vacuum must be very strongly correlated,
in an ``instanton liquid", not an ideal gas. 

Therefore, in this work we will first study the $correlations$ of different kind of the
instanton-dyons, using previously generated ensemble of configurations, for 
 $N_c=2$ and $N_f=2$ massless QCD. As we will show, the $\eta'$ phenomenon,
shifted from the correlator of the topological charges $<Q(x)Q(y)>$ to the correlator of 
$L \bar{L}$ dyons, is still there, and provides the strongest correlations of all channels. 

The last part of the paper is devoted to study of the $fluctuations$ of certain {\em global charges},
the topological, magnetic and electric ones, contained in a $fraction$
of the total system. Again, we will present the results from our ensemble of the generated
instanton-dyon
configurations. While those can be seen as just an integrated version of the same integrated correlation functions, the motivation to discuss those comes
from possible connection between our results and those obtained on the lattice.
Perhaps determination of global charges is technically a much easier task, compared to
complete identification of the instanton-dyons in the lattice configurations.

\section{The  correlations in the instanton-dyon ensemble } \label{sec_corr}
\subsection{The ensemble}

For simplicity, we discuss
 the simplest gauge group SU(2), the generalization
to SU(3) and other groups is straightforward. 
%their quantum numbers are in the table. 
For this work we use 
 previously generated ensemble of configurations with $64$ dyons, for 
 $N_c=2$ and $N_f=2$ massless QCD. These ensembles correspond to ten values
 of the temperature, $T/T_c=0.66, 0.72, 0.78, 0.85, 0.92, 1.0, 1.09, 1.18, 1.28, 1.44 $. Note that some of them are below and some above the
 deconfinement-chiral transition, which in this theory is of a weak crossover type.

 The correlation functions to be discussed below are defined and normalized as follows.
 We scan our stored configurations of the instanton-dyon ensembles and 
 put distances between the corresponding pairs into a file, which is then  histogramed. 
 To eliminate geometrical  factors, we normalize the resulting histogram to its version
 obtained with high-statistics set of randomly placed dyons. All of the resulting correlation
 functions approach unit value at large distances.

\subsection{$L\bar{L}$, $M\bar{M}$ and  $ML$ correlations}  \label{sec_LLbar}

The ``twisted" $L$-type dyons are the ones which have normalizable
 zero modes of the Dirac operator for  anti-periodic  quark fields on the Matsubara circle (that is, the physical fermions).  
This channel is therefore most interesting, because it is directly related with the effective $U(1)_a$ breaking and the
``$\eta'$ meson exchanges". 

\begin{figure}[htbp]
\begin{center}
\includegraphics[width=6.5cm]{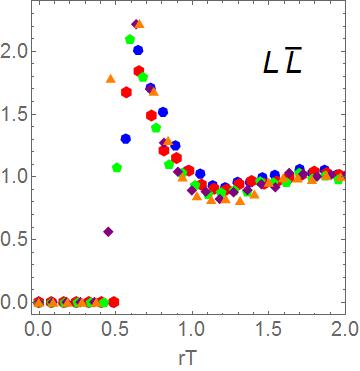}
\caption{(Color online) The correlation of  $L \bar{L}$ pair versus the distance $rT$. Here, and in all other plots, the points correspond to 
the following temperatures $T/T_c(color, polygon)=0.66 (blue, 7)$, $0.78 (red,6)$, $0.92 (green,5)$ ,  $1.09 (purple,4)$, $1.28 (orange, 3)$, in units of  $T_c$. Triangle is the highest temperature. 
}
\label{fig_LLbar}
\end{center}
\end{figure}

Our correlation function for interacting ensemble, normalized to that without interactions,
is shown in Fig.  \ref{fig_LLbar}. One can see no counts at small distances,
which is simply a manifestation of the presence of the ``repulsive core": the $L\bar{L}$
configurations below certain distance simply do not exist, see \cite{Larsen:2014yya}.

The next structure observed is a large positive correlation. It is induced by
both  the classical attraction in the  $L\bar{L}$ channel and the quark exchanges.
While the correlation effect is very strong, it also rapidly disappear with the distance,
more so than other correlation functions to be discussed. 
Its large-distance behavior should correspond to the effective $\eta'$ exchanges.
The fit to  screened on-sphere propagator (discussed in appendix B)
gives large mass $m/T_c \sim 3$, which is however only a half of what it is at $T=0$, in
the QCD vacuum, which is $m_{\eta'}/T_c=958/155\approx 6$.
Perhaps this difference show a partial $U(1)_a$ restoration, from   $T=0$ to $T\sim T_c$.

Note further a small dip below one seen at distance around $1.2$ in Fig. \ref{fig_LLbar}.
Oscillating correlation function, with a decreasing amplitude, are a clear sign of strongly correlated liquids.
 This effect is
 similar but much weaker than what one would see in a crystal.  At this time we are not sure what manybody structure has formed. 

\begin{figure}[htbp]
\begin{center}
\includegraphics[width=6.5cm]{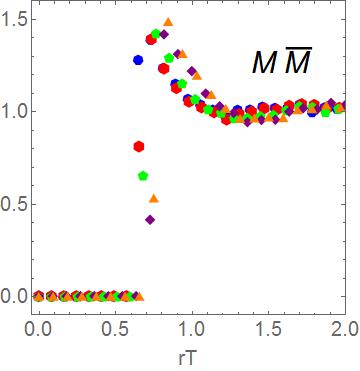}
\caption{(Color online) The correlation of  $M \bar{M}$ pair versus the distance $rT$. The different points corresponds to different temperatures as explained in Fig. \ref{fig_LLbar}. }
\label{fig_MMbar}
\end{center}
\end{figure}

%\subsection{ correlations}  \label{sec_MMbar}
The  correlation function for the $M\bar{M}$ dyon channel 
is shown in Fig. \ref{fig_MMbar}. Note that this channel has periodic rather than antiperiodic quark zero modes. This means that, unlike the previous case, the $M$-type dyons 
do not become 't Hooft effective vertices, and their correlator
does not include any quark exchanges. It indeed displays 
a similar core, but much 
smaller correlations. Those are 
induced by the classical (leading order) gluonic attraction studied in \cite{Larsen:2014yya}. 
Their large-distance asymptotics should thus provide a combination of electric and 
magnetic screening masses, which are $m/T_c\sim 2$.  

Note that unlike the $L\bar{L}$ correlator, the  $M\bar{M}$ one shows a systematic temperature dependence. The highest $T$, shown by triangles, show the strongest 
correlations. This happens because the classical (leading order) interaction
is $\sim 1/g^2(T)\sim log(T)$ gets stronger at high $T$. Indeed, the object discussed are
``magnetic", thus this unusual direction of running of their effective correlations.

\begin{figure}[htbp]
\begin{center}
\includegraphics[width=6.5cm]{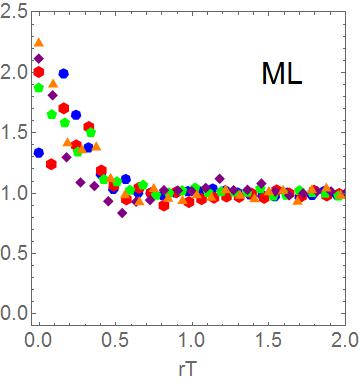}
\caption{(Color online) The correlation of  $ML$ pair versus the distance $rT$. The different points corresponds to different temperatures as explained in Fig. \ref{fig_LLbar}. }
\label{fig_ML}
\end{center}
\end{figure}

The last dyon channel we study is the
$ML$ one, corresponding  to the instanton-forming pair. It is 
the so called Bogomolny-Prasad-Sommerfeld (BPS) protected channel,
in which the  classical (leading order) interaction between the dyons vanishes. The 
correlation function shown in Fig. \ref{fig_ML} does not have the classical core,
and is concentrated at significantly smaller distances than other two: perhaps
indicating the instanton formation.
 The
observed effect is due to the
one-loop interaction studied by Diakonov et al \cite{Diakonov:2004jn}
and implemented in our partition function. 

%\subsection{ correlations}  \label{sec_ML}

\section{The fluctuations} \label{sec_fluct}

\subsection{The setting}
 Lattice gauge theories are traditionally defined on a 4-dimensional torus, by imposing periodic boundary conditions for all 4 coordinates. The 
 same geometry has been used for the  instanton liquid simulations. 
Recent studies of the instanton-dyon ensemble have been done using the
 $S^3$ sphere. The global charges -- e.g. magnetic $M$ and electric $E$ -- are fixed by the amount of dyons, and thus cannot fluctuate.

\begin{figure}[htbp]
\begin{center}
\includegraphics[width=6.5cm]{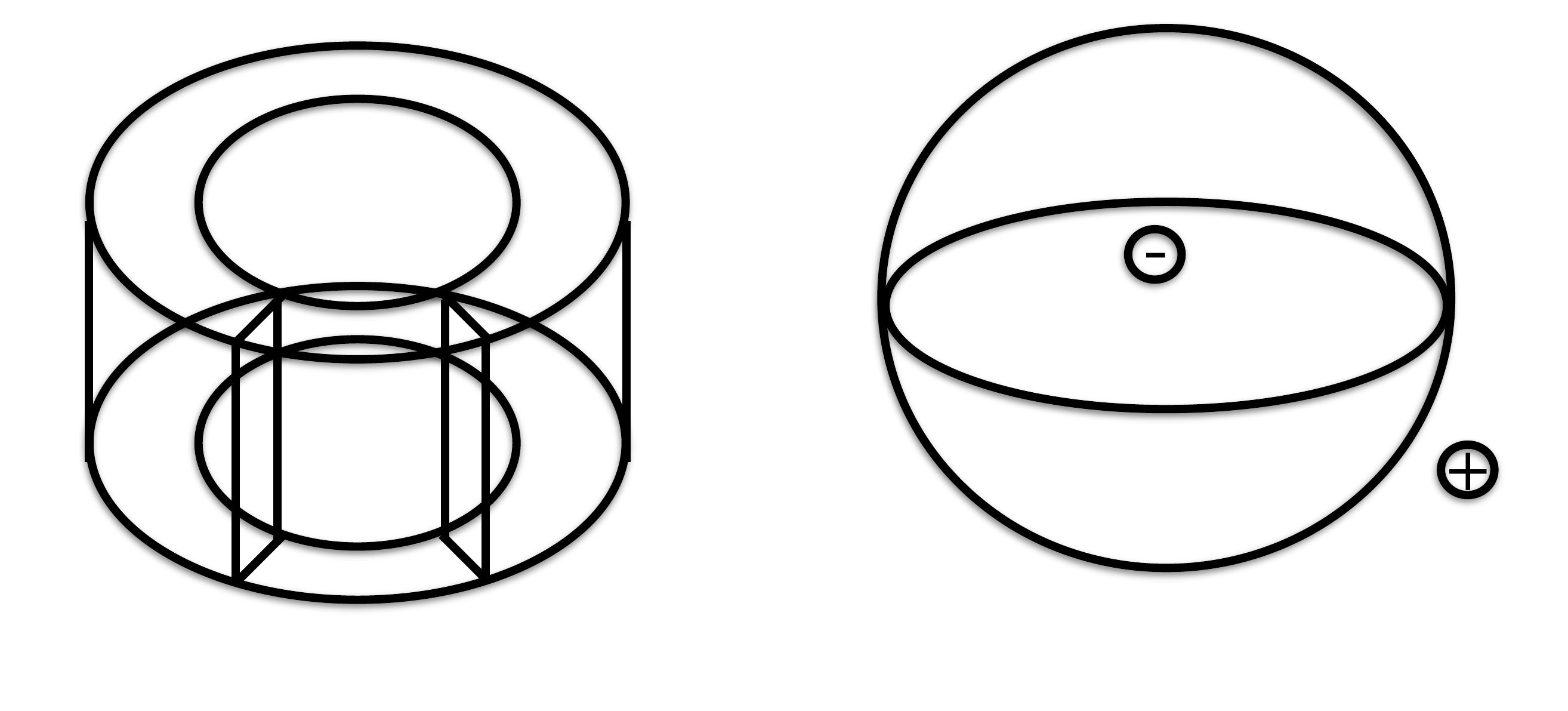}
\caption{(left)  A sketch of the torus with the subvolumes cut out; (right) 
The subvolumes of the $S^3$ sphere are, after a stereographic projection, the interior and exterior of the $S^2$ sphere as shown. The indicated $\pm$ charges can now be separated as shown. }
\label{fig_sketch}
\end{center}
\end{figure}

\begin{figure}[h]
\begin{center}
\includegraphics[width=8cm]{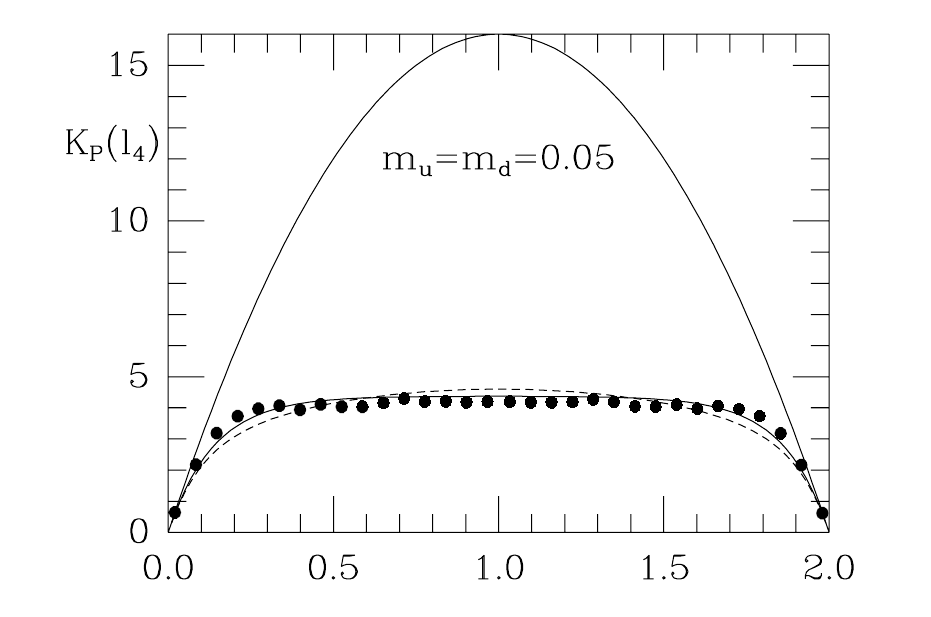}
\caption{An example of the susceptibility in sub-box $\chi_{sublat}$ as a function of the fraction of the total box, $x/L$, from  \cite{Shuryak:1994rr}.
The thin parabolic line corresponds to randomly placed instantons and antiinstantons, the dots are for the interacting instanton liquid. Strong screening of the topological charge in this model is evident.
Thin lines show different fits, from which the value of $m_{\eta'}$ was extracted.
}
\label{fig_screening}
\end{center}
\end{figure}

Cutting the torus by two planes, as shown in Fig. \ref{fig_sketch}(left), one can obtain  variable  $subvolumes$ $V_4$. Note that those have  $constant$ area $A_3$.
Therefore, the result of studies of the fluctuation of $Q$ in the instanton ensemble with light fermions,  reproduced in
Fig. \ref{fig_screening}, at large volume becomes horizontal (volume-independent).
 If the volume in question, e.g. the 4-d torus used in lattice  or
 the $S^3$ used in the instanton-dyon simulations, has a boundary, like in sub-volumes to be
discussed, the fluctuations occur near it, basically in the volume
\be V_{fluct}\sim A_{boundary} R_{screening} \ee

The setting of the instanton-dyon simulations we will study is different. The $S^3$
sphere, imbedded into 4-d space, can be cut into two parts by a $single$ plane.
It is convenient to think about the stereographic projection 
of this sphere back to the 3-d space. The  resulting boundary is the $S^2$ sphere, with
the 3-d subvolumes being its interior and exterior  space, as shown
in Fig. \ref{fig_sketch}(right).  We remind that the volumes of both subvolumes
are finite, see details in the appendix. When the radius of the 
boundary grows, its area reaches a maximum, at the angle $\psi=\pi/2$, and then decreases,
due to the $r$-dependent factor induced by the  stereographic projection.

\subsection{The fluctuating charges}

One can either study fluctuations of each dyon
type, or of their particular combinations. We propose instead to focus
on the fluctuations of the following global quantum numbers, the topological charge $Q$, the magnetic and electric
charges $M$ and $E$, and the  action $S$
\be \chi_Q =<Q^2> \nonumber \\
 \chi_M =<M^2> \nonumber \\
 \chi_E =<E^2> -<E>^2\\
\chi_S =<S^2> -<S>^2 \nonumber \ee
Note that  the last two have a non-zero VEV, which needs to be subtracted.
The average  magnetic charge $<M>=0$ because dyon ensembles have equal number of dyons and anti-dyons. The electric charge is different, because in general $n_M\neq n_L$.

The former one is the well known topological susceptibility, long studied on the lattice. 
Its usual definition includes division by the 4-volume $V_4$ and the limit $V_4\rightarrow \infty$, which eliminates the dependence on the volume shape. As it was already
emphasized in \cite{Shuryak:1994rr}, not only is this limit unnecessary, the study
of the fluctuations dependence on the volume shape and size reveals such valuable information as {\em the screening lengths} for the corresponding charges.
For $Q$ it is known as the $\eta'$ mass, and for $M,E$ as the magnetic and electric screening
masses $m_M,m_E$ -- all three subject for separate lattice measurements. 

Let us further remind that in QCD-like theories with light quarks the  topological susceptibility
vanishes in the chiral limit $m\rightarrow 0$ and is
suppressed by the powers of the nonzero quark masses, $O(m)$ at $T<T_c$ and $O(m^{N_f})$ at  $T>T_c$. The latter is just the  suppression due to 't Hooft zero modes.

At high $T > 2T_c$ the VEV of the Polyakov line $<P>\approx 1$ and the
 nontrivial holonomy fields are practically absent. That means the topological objects
 are the $instantons$ (or, more precisely, their finite-$T$ version known as calorons).
 Since at such $T$ the instanton actions are large and density exponentially small, 
 one expects, and indeed observes on the lattice, that the topological ensemble is
 represented by the Dilute Instanton Ensemble (DIE).  Diluteness leads to the Poisson
 statistics of fluctuations, and thus $\chi_Q$  basically gives us  the instanton density.
  
 At lower $T < 300-400\, MeV$, the VEV of the Polyakov line takes some values between 1 and 0, and the appropriate ensemble should be described in terms of the 
  instanton-dyons. While all of them have the topological charge  
  $Q$, only ``twisted" $L$-kind have fermionic zero modes. Therefore, fluctuations of $Q$
  and fermion mass suppression should in principle become decoupled. In particularly,
  in the chiral limit there should be a non-zero $\chi_Q$ due to $M$-type dyons.
  
   \begin{figure}[h]
\begin{center}
\includegraphics[width=8.cm]{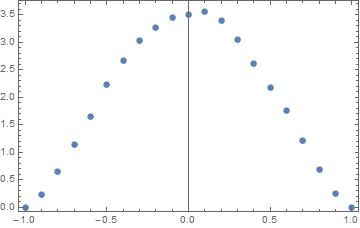}
\caption{The mean topological charge squared $<Q^2>$ in a subvolume, versus 
the cut parameter $cos(\psi_{cut})$.}
\label{fig_qq_unnormalized}
\end{center}
\end{figure}
  
  \subsection{Fluctuations in the instanton-dyon ensemble}
  
  In Fig. \ref{fig_qq_unnormalized} we show a typical 
   histogram, displaying fluctuation of the topological charge $<Q^2>$ 
   as a function of the size of the subvolume. It has a characteristic symmetric  shape,
   because fluctuations in a small volume and in a complementary volume which includes the whole sphere
   without a small part, are identical. 
   
   In order to understand them better, one needs to normalize such histograms to 
  a distribution calculated for uncorrelated dyons. The uncorrelated case is only volume dependent and one can therefore write the probability for finding $n$ out of $N$ charges in a sub-volume $V$ of the total volume $V_0$, as
\begin{eqnarray}
P(V,n) &=& \frac{N!}{n! (N-n)!}\left(\frac{V}{V_0}\right)^n\left(1-\frac{V}{V_0}\right)^{N-n}
\end{eqnarray}
Which gives the fluctuation when summed over the charge squared for all $n$. 
 
\begin{figure}[h!]
\begin{center}
\includegraphics[width=4.0cm]{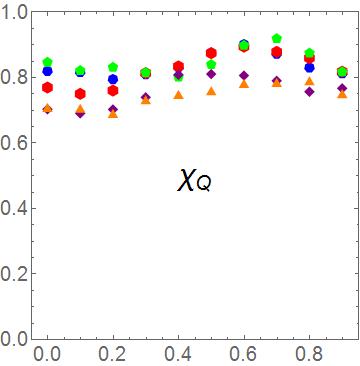}
\includegraphics[width=4.0cm]{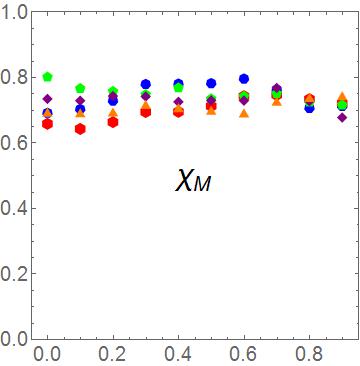} \\
\includegraphics[width=4.0cm]{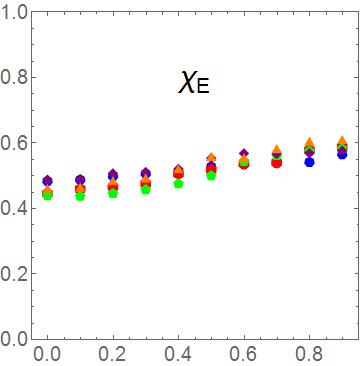} 
\includegraphics[width=4.0cm]{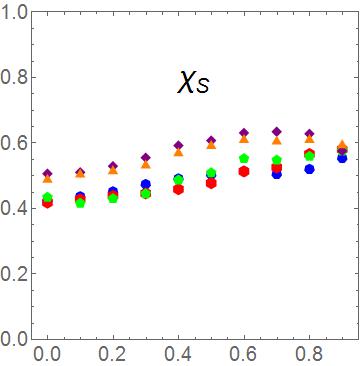}
\caption{(Color online) The normalized fluctuations of the topological charge $Q$, the magnetic charge $M$,
the electric charge $E$ and the action $S$, as a function of subvolume cut $cos(\psi_{cut})$. 
Because of symmetry of the distributions, only one half of it is shown.
The l.h.s., $cos(\psi_{cut})=0$ corresponds to cutting the sphere into two equal halfs, the r.h.s.
at  $cos(\psi_{cut})\rightarrow 1$
corresponds to cutting off a very small part. The different points corresponds to different temperatures as explained in Fig. \ref{fig_LLbar}.
}
\label{fig_fluc}
\end{center}
\end{figure}

    The resulting plots are shown in 
    Fig. \ref{fig_fluc}. Note, that if the ensemble be an ideal gas, 
     by definition all plots should show values equal to one.
 
% \begin{widetext}
%\end{widetext}

   Few comments on those are in order: 
   \\(i) The 
 fluctuations of all charges are very similar. They all seem to be
 of the ``volume" kind, since the plots normalized to random ensemble all look rather flat. 
 \\ (ii)   The absolute value of the fluctuations is however not one, but smaller. This 
   reflects presence of an attraction between opposite charges, resulting in
   formation of ``neutral clusters", pairs etc.
   \\(iii)   Like for the correlations discussed above,
   we observe only very small temperature dependence of the fluctuations.
  \\(iv)   
     There are visible deviations from constant value 
   of the normalized fluctuations at the ``wings" of the distributions.
   Those appear when the sub-volume dimension
   becomes comparable to the micro correlation lengths of the objects themselves.
   \\(v)
   Some combinations of dyon interactions have a hard core in the ensemble. This cuts out possible small fluctuations, and is at least partially responsible for the values far off one at small volumes, ie. x-axis close to 1.
   
    We do not try fitting those deviations for the values of the screening masses,
  as those were much better displayed
  by the correlation measurements reported above.

\section{Summary and discussion}

Short-range correlations between $L\bar{L}$ dyons is the channel corresponding
to the $<QQ>$ correlator in the QCD vacuum, in the sense that both are related to 
quark-induced forces and large
$\eta'$ mass. The main conclusions one can get from our discussion of this
correlation function in section \ref{sec_LLbar} are: (i)  like in vacuum, this channel
shows the largest screening mass; (ii) the screening effect 
in studied $T$ range is basically $T$-independent.

This is to be compared to the long-distance correlations associated with
chiral symmetry breaking. For $T<T_c$  we know that there appear 
very long ``topological clusters", leading to
very small Dirac eigenvalues, which are absent at $T>T_c$. So, two 
different approaches to the topology correlations are clearly complementary.

The  $M\bar{M}$ dyon channel does not have quark exchanges. It display smaller correlations
induced by classical (leading order) attraction studied in \cite{Larsen:2014yya}. 
The $ML$ dyon channel corresponds to the instanton-forming pair. It is 
the so called Bogomolny-Prasad-Sommerfeld (BPS) protected channel,
in which the  classical (leading order) interaction between the dyons vanishes. 

The second part of the paper is devoted to $fluctuations$ of certain charges in subvolumes.
In principle, those should be  sensitive to corresponding screening lengths, provided
those are comparable with the actual dimensions of these  subvolumes.
However 
the data presented in section \ref{sec_fluct} do $not$ show any clear shape difference between  the interacting ensemble of dyons and the random one:  perhaps all of them only depend on the size of the subvolume. 

These results are by no means trivial, they can and should be used
in order to determine on the lattice where the transition from the instanton gas to
the instanton-dyon plasma takes place. Recall that the instanton ensemble has 
no  fluctuations of the magnetic charge at all, 
while the  instanton-dyon plasma predicts certain relations between 
topological and magnetic charges, as shown above.
The instanton ensemble in the chiral limit $m_q\rightarrow 0$ has
no $\chi$ proportional to the volume, since single instantons in this limit disappear. 
The  instanton-dyon ensemble, on the other hand, even in the chiral limit,
 has non-zero $\chi$ fluctuations due to $M$-dyons.

\appendix 
\section{The instanton-dyon ensemble}
This section contain brief reminder of the model used. 
The so called instanton-dyons are instanton constituents.
Their sizes and actions depend on one crucial parameter, the so called {\em holonomy value} $\nu(T)$, related to the mean value of the Polyakov line. At high $T > 2T_c$ the VEV of the Polyakov line $<P>\approx 1$ and the holonomy disappears, $\nu\approx 0$. At
 $T < T_c$ $<P>\approx 0$ and the holonomy takes the so called ``confining value".
In the considered case of the $SU(2)$ color   $\nu\approx 1/2$.

The two-color theory has  4 instanton-dyon types,
$M,L,\bar{M},\bar{L}$. $M$ and $L$ are selfdual, the other anti-selfdual.
The actions of those, and their interactions, have been defined semiclassically. 
Certain number of the dyons -- in the ensemble used 64 of them -- are put on the
3-dimensional sphere $S^3$. Standard Metropolis algorithm and ``integration back"
method are used to calculate the free energy.
The value of the parameters -- such as holonomy value $\nu$ and the
 densities of different types, $n_M \neq n_L$ -- are defined from the minimum of the free energy, and thus are temperature dependent.
 
 Briefly, about geometry of the $S^3$ sphere. Its volume is $V_3=2\pi^2 R^3$, where $R$
 is the radius. Since the dyon density decreases with temperature, and we keep
 their number fixed, this radius is also increasing with $T$. 
 
 We work with locations on the $S^3$ sphere using 4-dimensional unit vectors.
% The distance squared between them is simply $R^2 (\vec n_1 \cdot \vec n_2)$
 Standard three standard polar angles $\psi,\theta,\phi$ can be used to define coordinates
 $$  \vec n=[ \sin(\psi] \sin(\theta)\sin(\phi), 
 \sin(\psi) \sin(\theta) \cos(\phi),  $$ \be 
 \sin(\psi) \cos(\theta), \cos(\psi) ]\ee
The corresponding 
 metrics is
\be dl^2/R^2 =d\psi^2+sin^2\psi d\theta^2 + sin^2\psi sin^2\theta d\phi^2, \ee
and the volume element reads \be dV=\sin^2(\psi)\sin(\theta) d\psi d\theta d\phi \ee
Subvolumes, used in the fluctuation section, are defined via the boundary plane $\psi=\psi_{cut}$.

The points on the sphere can be mapped onto points on its equatorial plane $X^i,i=1,2,3$ 
via standard stereographic projection. The subvolumes correspond to the interior and exterior of a $S^2$ sphere, with the radius defined by $\psi_{cut}$. The projected volume element 
on the  plane $X^i,i=1,2,3$ is given by a version of Riemann formula
\be dV= ({2 \over 1+ r^2/R^2})^3 4\pi r^2 dr \ee
  The first factor makes the integral to $r=\infty$ convergent, and equal to that of the
  original sphere.

\section{Screening in the infinite volume and on the $S^3$ sphere} 

%  \section{The relation between the screening lengths and the shape/size of the volume}

%Now for dyons: it is 3d and will be used for a $S^3$ sphere.

General expressions for the point-to-point 3d correlator of charges, e.g. the topological charge $Q$, takes the standard Yukawa form with a screening  mass (inverse radius) $m$
\be K(|x-y|)=<Q(x) Q(y)>= \\
\delta(x-y) - {1 \over 4 \pi |x-y|} exp(- {|x-y| m}) \nonumber \ee
in which the first term is say a positive charge and the second is its
``compensated" negative charge. The normalization corresponds to the integral
over either $x$ or $y$ to vanish, $<Q>=0$.
This feature is also well known in  the $q \rightarrow 0$ limit of
its Fourier transform 
\be \tilde K(q)= 1- {1 \over q^2 R^2 +1}=  {q^2 R^2 \over q^2 R^2 +1} \ee
In flat space fluctuations in a subvolume $V$ can be described by the double integral
 \be I_V=\int_{x,y\in V} d^3x d^3y K(|x-y|) \ee

%The cut of the $S^3$ sphere by a plane separate two sub-volumes. Their common boundary is 
% the $S^2$ sphere, but the volume is not just that inside the sphere 
% because the underlying metric is not flat. The volumes are $V(\chi_{cut})= ??? $
% where coordinate on the 3-sphere are $\chi,\theta,\phi$. 
%The data needs to be compared to the following quantity
%Returning to the Fourier form of the function $K$, one separates $x$ and $y$ integrals and
%\be I_V=\int {d^3q \over (2\pi)^3} |V_q|^2 {q^2 R^2 \over q^2 R^2 +1} \ee
%where \be V_q=\int_{x\in V} \sqrt{g} d^3x e^{i\vec q \vec x} \ee

On  the 3-dimensional sphere $S^3$ one needs to find the analog to the Yukawa potential.
%geometry, as an example of very symmetric manifold without a boundary. 
%The points are defined via coordinates $x^i,i=1,4$ satisfying $\sum_{i=1}^4 (x^i)^2=R^2$.
%The radius of the sphere $R$ is related to the volume of it $V_{S3}=2\pi^2 R^3$,
%tuned to have the density of the objects we would like to simulate.
%
%
%Introducing the usual spherical coordinates with three angles $\psi,\theta,\phi$
 The screened  Coulomb potential
of a charge, placed on the north pole of the $S^3$ sphere $\psi=0$, 
can be found from  the $\psi$-part of the Laplacian, which reads
 \be {1 \over \sqrt{g}} \partial_\mu (\sqrt{g} g^{\mu\nu} \partial_\nu f ) =f''(\psi)+{2 \over tan(\psi)} f'(\psi)\ee
%Recall that in flat space the screened Coulomb potential is solution of the Laplace eqn with 
Adding 
the mass term, one finds the needed analogue of
the  Yukawa potential to be  
\be f(\psi)= { exp[-\sqrt{-1 + m^2 R^2} \psi]  \over  4\pi sin(\psi)} \ee
Note the presence of the second singularity at the south pole $\psi=\pi$, which is
however exponentially suppressed by $exp(- m R \pi) $ and disappear for large macroscopic spheres $mR \gg 1$. 

%The points on the sphere can be mapped onto points on the plane $X^i,i=1,2,3$ 
%via standard stereographic projection
%\be X^i \ee

%The fluctuations of charges in ``sliced" part of the  $S^3$ sphere, defined by a condition
%$\psi<\psi_{cut}$, is the integrated two-particle correlation function
%\be \int_{\psi_1<\psi_{cut}}  d\Omega_1 \int_{\psi_2<\psi_{cut}}d\Omega_2 f(\Delta_{12})\ee
%where the screening function $f$ defined above is a function of the distance between the two points, defined in our 4-d vector representation by the scalar product 
%\be \Delta_{12} =R \cdot arccos(
%n_1^\mu n_2^\mu) \ee  
%
%
%\section{$S^3$ stereographic projection and related formulae }

%%%%%%%%%%%%%%%%%%%%%%%%%%%%%%%%%%%%%%


\begin{thebibliography}{99}

%\cite{Belavin:1975fg}
\bibitem{Belavin:1975fg} 
  A.~A.~Belavin, A.~M.~Polyakov, A.~S.~Schwartz and Y.~S.~Tyupkin,
  %``Pseudoparticle Solutions of the Yang-Mills Equations,''
  Phys.\ Lett.\  {\bf 59B}, 85 (1975).
  doi:10.1016/0370-2693(75)90163-X
  %%CITATION = doi:10.1016/0370-2693(75)90163-X;%%
  %2612 citations counted in INSPIRE as of 03 Mar 2017

%\cite{Schafer:1996wv}
\bibitem{Schafer:1996wv} 
  T.~Schafer and E.~V.~Shuryak,
  %``Instantons in QCD,''
  Rev.\ Mod.\ Phys.\  {\bf 70}, 323 (1998)
  doi:10.1103/RevModPhys.70.323
  [hep-ph/9610451].
  %%CITATION = doi:10.1103/RevModPhys.70.323;%%
  %1196 citations counted in INSPIRE as of 03 Mar 2017


%\cite{Kraan:1998sn}
\bibitem{Kraan:1998sn} 
  T.~C.~Kraan and P.~van Baal,
  %``Monopole constituents inside SU(n) calorons,''
  Phys.\ Lett.\ B {\bf 435}, 389 (1998)
  doi:10.1016/S0370-2693(98)00799-0
  [hep-th/9806034].
  %%CITATION = doi:10.1016/S0370-2693(98)00799-0;%%
  %210 citations counted in INSPIRE as of 03 Mar 2017


%\cite{Lee:1998bb}
\bibitem{Lee:1998bb} 
  K.~M.~Lee and C.~h.~Lu,
  %``SU(2) calorons and magnetic monopoles,''
  Phys.\ Rev.\ D {\bf 58}, 025011 (1998)
  doi:10.1103/PhysRevD.58.025011
  [hep-th/9802108].
  %%CITATION = doi:10.1103/PhysRevD.58.025011;%%
  %255 citations counted in INSPIRE as of 03 Mar 2017


%\cite{Davies:1999uw}
\bibitem{Davies:1999uw} 
  N.~M.~Davies, T.~J.~Hollowood, V.~V.~Khoze and M.~P.~Mattis,
  %``Gluino condensate and magnetic monopoles in supersymmetric gluodynamics,''
  Nucl.\ Phys.\ B {\bf 559}, 123 (1999)
  doi:10.1016/S0550-3213(99)00434-4
  [hep-th/9905015].
  %%CITATION = doi:10.1016/S0550-3213(99)00434-4;%%
  %168 citations counted in INSPIRE as of 03 Mar 2017


%\cite{Poppitz:2011wy}
\bibitem{Poppitz:2011wy} 
  E.~Poppitz and M.~Unsal,
  %``Seiberg-Witten and 'Polyakov-like' magnetic bion confinements are continuously connected,''
  JHEP {\bf 1107}, 082 (2011)
  doi:10.1007/JHEP07(2011)082
  [arXiv:1105.3969 [hep-th]].
  %%CITATION = doi:10.1007/JHEP07(2011)082;%%
  %47 citations counted in INSPIRE as of 03 Mar 2017


%\cite{Poppitz:2012sw}
\bibitem{Poppitz:2012sw} 
  E.~Poppitz, T.~Schafer and M.~Unsal,
  %``Continuity, Deconfinement, and (Super) Yang-Mills Theory,''
  JHEP {\bf 1210}, 115 (2012)
  doi:10.1007/JHEP10(2012)115
  [arXiv:1205.0290 [hep-th]].
  %%CITATION = doi:10.1007/JHEP10(2012)115;%%
  %59 citations counted in INSPIRE as of 03 Mar 2017


%\cite{Liu:2015ufa}
\bibitem{Liu:2015ufa} 
  Y.~Liu, E.~Shuryak and I.~Zahed,
  %``Confining dyon-antidyon Coulomb liquid model. I.,''
  Phys.\ Rev.\ D {\bf 92}, no. 8, 085006 (2015)
  doi:10.1103/PhysRevD.92.085006
  [arXiv:1503.03058 [hep-ph]].
  %%CITATION = doi:10.1103/PhysRevD.92.085006;%%
  %25 citations counted in INSPIRE as of 03 Mar 2017


%\cite{Liu:2015jsa}
\bibitem{Liu:2015jsa} 
  Y.~Liu, E.~Shuryak and I.~Zahed,
  %``Light quarks in the screened dyon-antidyon Coulomb liquid model. II.,''
  Phys.\ Rev.\ D {\bf 92}, no. 8, 085007 (2015)
  doi:10.1103/PhysRevD.92.085007
  [arXiv:1503.09148 [hep-ph]].
  %%CITATION = doi:10.1103/PhysRevD.92.085007;%%
  %20 citations counted in INSPIRE as of 03 Mar 2017


%\cite{Liu:2016thw}
\bibitem{Liu:2016thw} 
  Y.~Liu, E.~Shuryak and I.~Zahed,
  %``The Instanton-Dyon Liquid Model III: Finite Chemical Potential,''
  Phys.\ Rev.\ D {\bf 94}, no. 10, 105011 (2016)
  doi:10.1103/PhysRevD.94.105011
  [arXiv:1606.07009 [hep-ph]].
  %%CITATION = doi:10.1103/PhysRevD.94.105011;%%
  %2 citations counted in INSPIRE as of 03 Mar 2017


%\cite{Liu:2016mrk}
\bibitem{Liu:2016mrk} 
  Y.~Liu, E.~Shuryak and I.~Zahed,
  %``Light Adjoint Quarks in the Instanton-Dyon Liquid Model IV,''
  Phys.\ Rev.\ D {\bf 94}, no. 10, 105012 (2016)
  doi:10.1103/PhysRevD.94.105012
  [arXiv:1605.07584 [hep-ph]].
  %%CITATION = doi:10.1103/PhysRevD.94.105012;%%
  %2 citations counted in INSPIRE as of 03 Mar 2017


%\cite{Liu:2016yij}
\bibitem{Liu:2016yij} 
  Y.~Liu, E.~Shuryak and I.~Zahed,
  %``The Instanton-Dyon Liquid Model V: Twisted Light Quarks,''
  Phys.\ Rev.\ D {\bf 94}, no. 10, 105013 (2016)
  doi:10.1103/PhysRevD.94.105013
  [arXiv:1606.02996 [hep-ph]].
  %%CITATION = doi:10.1103/PhysRevD.94.105013;%%
  %2 citations counted in INSPIRE as of 03 Mar 2017


%\cite{Faccioli:2013ja}
\bibitem{Faccioli:2013ja} 
  P.~Faccioli and E.~Shuryak,
  %``QCD topology at finite temperature: Statistical mechanics of self-dual dyons,''
  Phys.\ Rev.\ D {\bf 87}, no. 7, 074009 (2013)
  doi:10.1103/PhysRevD.87.074009
  [arXiv:1301.2523 [hep-ph]].
  %%CITATION = doi:10.1103/PhysRevD.87.074009;%%
  %27 citations counted in INSPIRE as of 03 Mar 2017


%\cite{Larsen:2014yya}
\bibitem{Larsen:2014yya} 
  R.~Larsen and E.~Shuryak,
  %``Classical interactions of the instanton-dyons with antidyons,''
  Nucl.\ Phys.\ A {\bf 950}, 110 (2016)
  doi:10.1016/j.nuclphysa.2016.03.013
  [arXiv:1408.6563 [hep-ph]].
  %%CITATION = doi:10.1016/j.nuclphysa.2016.03.013;%%
  %18 citations counted in INSPIRE as of 03 Mar 2017


%\cite{Larsen:2015vaa}
\bibitem{Larsen:2015vaa} 
  R.~Larsen and E.~Shuryak,
  %``Interacting ensemble of the instanton-dyons and the deconfinement phase transition in the SU(2) gauge theory,''
  Phys.\ Rev.\ D {\bf 92}, no. 9, 094022 (2015)
  doi:10.1103/PhysRevD.92.094022
  [arXiv:1504.03341 [hep-ph]].
  %%CITATION = doi:10.1103/PhysRevD.92.094022;%%
  %14 citations counted in INSPIRE as of 03 Mar 2017


%\cite{Larsen:2015tso}
\bibitem{Larsen:2015tso} 
  R.~Larsen and E.~Shuryak,
  %``Instanton-dyon Ensemble with two Dynamical Quarks: the Chiral Symmetry Breaking,''
  Phys.\ Rev.\ D {\bf 93}, no. 5, 054029 (2016)
  doi:10.1103/PhysRevD.93.054029
  [arXiv:1511.02237 [hep-ph]].
  %%CITATION = doi:10.1103/PhysRevD.93.054029;%%
  %11 citations counted in INSPIRE as of 03 Mar 2017


%\cite{Larsen:2016fvs}
\bibitem{Larsen:2016fvs} 
  R.~Larsen and E.~Shuryak,
  %``Instanton-dyon ensembles with quarks with modified boundary conditions,''
  Phys.\ Rev.\ D {\bf 94}, no. 9, 094009 (2016)
  doi:10.1103/PhysRevD.94.094009
  [arXiv:1605.07474 [hep-ph]].
  %%CITATION = doi:10.1103/PhysRevD.94.094009;%%
  %3 citations counted in INSPIRE as of 03 Mar 2017


%\cite{Shuryak:2016vow}
\bibitem{Shuryak:2016vow} 
  E.~Shuryak,
  %``Recent progress in understanding deconfinement and chiral restoration phase transitions,''
  arXiv:1610.08789 [nucl-th].
  %%CITATION = ARXIV:1610.08789;%%
  %1 citations counted in INSPIRE as of 03 Mar 2017

\bibitem{Bonati}  C. Bonati, et al., JHEP 03 (2016) 155. arxiv 1612.06269

\bibitem{Borsanyi} S. Borsanyi, et al., Phys. Lett. B752 (2016) 175Ð181.
%\cite{Borsanyi:2016ksw}
%\bibitem{Borsanyi:2016ksw} 
%  S.~Borsanyi {\it et al.},
  %``Calculation of the axion mass based on high-temperature lattice quantum chromodynamics,''
  Nature {\bf 539}, no. 7627, 69 (2016)
  doi:10.1038/nature20115
  [arXiv:1606.07494 [hep-lat]].
  %%CITATION = doi:10.1038/nature20115;%%
  %29 citations counted in INSPIRE as of 24 Jan 2017
  
%\cite{Petreczky:2016vrs}
\bibitem{Petreczky:2016vrs} 
  P.~Petreczky, H.~P.~Schadler and S.~Sharma,
  %``The topological susceptibility in finite temperature QCD and axion cosmology,''
  Phys.\ Lett.\ B {\bf 762}, 498 (2016)
  doi:10.1016/j.physletb.2016.09.063
  [arXiv:1606.03145 [hep-lat]].
  %%CITATION = doi:10.1016/j.physletb.2016.09.063;%%
  %13 citations counted in INSPIRE as of 24 Jan 2017



%\cite{Bornyakov:2015xao}
\bibitem{Bornyakov:2015xao} 
  V.~G.~Bornyakov, E.-M.~Ilgenfritz, B.~V.~Martemyanov and M.~MŸller-Preussker,
  %``Dyons near the transition temperature in lattice QCD,''
  Phys.\ Rev.\ D {\bf 93}, no. 7, 074508 (2016)
  doi:10.1103/PhysRevD.93.074508
  [arXiv:1512.03217 [hep-lat]].
  %%CITATION = doi:10.1103/PhysRevD.93.074508;%%
  %8 citations counted in INSPIRE as of 03 Mar 2017

%\cite{Shuryak:2017fkh}
\bibitem{Shuryak:2017fkh} 
  E.~Shuryak,
  %``Comments on the temperature dependence of the gauge topology,''
  arXiv:1701.08089 [hep-lat].
  %%CITATION = ARXIV:1701.08089;%%

%\cite{Shuryak:1994rr}
\bibitem{Shuryak:1994rr} 
  E.~V.~Shuryak and J.~J.~M.~Verbaarschot,
  %``Screening of the topological charge in a correlated instanton vacuum,''
  Phys.\ Rev.\ D {\bf 52}, 295 (1995)
  doi:10.1103/PhysRevD.52.295
  [hep-lat/9409020].
  %%CITATION = doi:10.1103/PhysRevD.52.295;%%
  %46 citations counted in INSPIRE as of 03 Mar 2017

%\cite{Diakonov:2004jn}
\bibitem{Diakonov:2004jn} 
  D.~Diakonov, N.~Gromov, V.~Petrov and S.~Slizovskiy,
  %``Quantum weights of dyons and of instantons with nontrivial holonomy,''
  Phys.\ Rev.\ D {\bf 70}, 036003 (2004)
  doi:10.1103/PhysRevD.70.036003
  [hep-th/0404042].
  %%CITATION = doi:10.1103/PhysRevD.70.036003;%%
  %113 citations counted in INSPIRE as of 03 Mar 2017


\end{thebibliography}
\end{document}